\documentclass[doublecol]{epl2}
\usepackage{graphicx}
\usepackage{amsmath,bbm,amssymb,amscd,bbold,mathbbol}

\newcommand{\be}{\begin{equation}}
\newcommand{\ee}{\end{equation}}
\newcommand{\bea}{\begin{eqnarray}}
\newcommand{\eea}{\end{eqnarray}}
\newcommand{\bra}[1]{\langle #1|}
\newcommand{\ket}[1]{|#1\rangle}

\title{Optimal Control of Superconducting N-level quantum systems}

\shorttitle{}

\author{H. Jirari\inst{1} \and F. W. J. Hekking\inst{1} \and O. Buisson\inst{2}}
\shortauthor{H. Jirari \etal}

\institute{
  \inst{1} LPMMC, C.N.R.S- Universit\'e Joseph Fourier,
    BP 166, 38042 Grenoble-cedex 9, France\\
  \inst{2} Institut N\'eel, C.N.R.S- Universit\'e Joseph Fourier,
    BP 166, 38042 Grenoble-cedex 9, France }


\abstract{We consider a current-biased dc SQUID in the presence of
an applied time-dependent bias current or magnetic flux. The phase dynamics of
such a Josephson device is equivalent to that of a quantum particle
trapped in a $1-$D anharmonic potential, subject to
external time-dependent control fields, {\it i.e.} a driven
multilevel quantum system. The problem of finding the required
time-dependent control field that will steer the system from a
given initial state to a desired final state at a specified final
time is formulated in the framework of optimal control theory.
Using the spectral filter technique, we show that the selected
optimal field which induces a coherent population transfer between
quantum states is represented by a carrier signal having a constant
frequency but which is time-varied both in amplitude and phase. The sensitivity of the
optimal solution to parameter perturbations is also addressed.}

\begin{document}
\maketitle

\section{Introduction} Superconducting circuits with Josephson
junctions have received a lot of attention recently as
promising candidates for scalable quantum
bits~\cite{makhlin01,wendin05}. An example of such a circuit is
the so-called phase qubit~\cite{martinis02,neeley08,hoskinson08}, which is based on a
current-biased Josephson junction. The phase dynamics of this
Josephson device is analogous to that of a quantum particle
trapped in a $1-$D anharmonic potential. Preparation and control
of the quantum states of the anharmonic well can be achieved by
applying time-dependent current pulses to the system. The device
can be used as a qubit when operated in the lowest two eigenstates
of the anharmonic well.

The energy levels beyond the lowest two can be addressed as
well. In particular, Rabi-like oscillations in the multilevel limit
have been observed with a current-biased dc-SQUID~\cite{cblaudo04}.
In the context of quantum information processing, the coupling
between the computational basis and the states of the
noncomputational subspace results in adverse effects on quantum
gate operations~\cite{fazio99}. However, there is no need to
restrict to only two energy levels. A generalization to qudits
({\it i.e.} systems with a single particle Hilbert space of dimension
$d>2$) has been proposed for quantum
computation~\cite{genovese07}. In this case, the quantum
information is encoded in higher-dimensional Hilbert spaces.

Quantum computation requires a precise and complete control of quantum systems. The
purpose of the present work is to apply optimal control theory to accurately transfer the
populations of qudit states present in a current-biased dc-SQUID. The theory of optimal
control is a well-developed field and finds numerous applications to the optimisation of
nonlinear and highly complex dynamic systems~\cite{ARTHURE}. In the quantum chemistry
context, optimal control was originally proposed by Rabitz and co-workers~\cite{shi88} as
a control scheme of reaction channels and was extensively used in various control
experiments. Optimal control theory provides a systematic and flexible formalism that can
be used in quantum computation to generate reliable and high precision quantum
dynamics~\cite{grace07}. A very recent application deals with the optimization of a NOT-gate for phase qubits~\cite{safaei_08}.

In the qudit case, population transfer can be
realised via coherent transitions between quantum states
interacting with the external control. In this letter we will use
optimal control theory to demonstrate the possibility of a
population transfer from the ground state to an arbitrary excited
state of a quantum N-level system. In general, correlation and
interference between the various pathways involved in the population
transfer process cannot be ignored and lead to a possibly
complicated dependence of the optimal control field on time which
may be difficult to implement experimentally. We therefore
restrict the frequency content of the control field
using the spectral filter technique~\cite{werschnik07,gross91}.
This enables us to find optimized control
fields that may be experimentally feasible.

\section{Model} 
\begin{figure}[!t]
\centerline{\includegraphics[width=1\columnwidth]{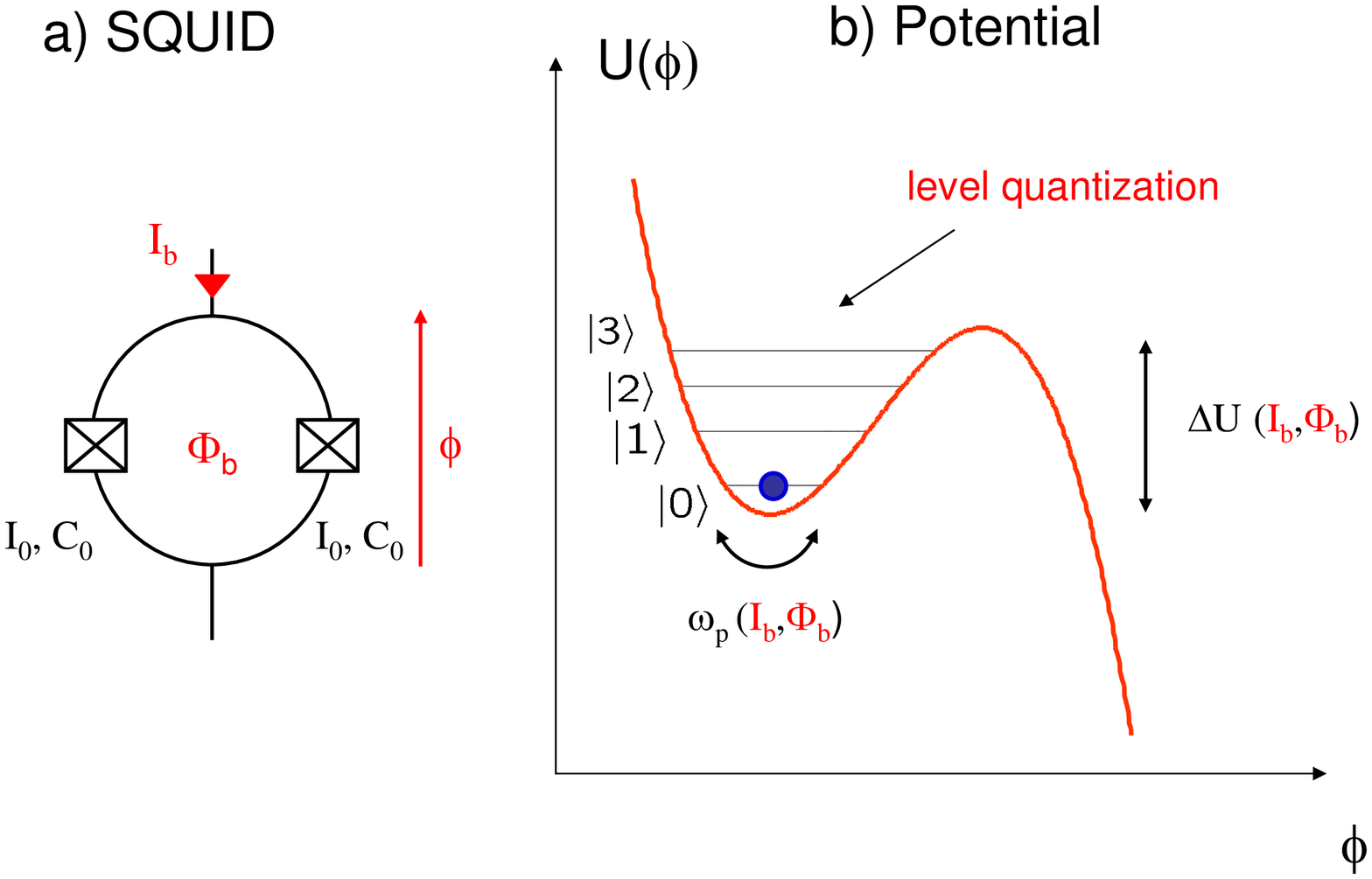}}
\caption{(a) Current-biased DC SQUID. (b) Phase-dependent anharmonic potential with quantized levels.}
\label{system}
\end{figure}

We consider a dc-SQUID, biased with a current $I_b$, consisting of two Josephson tunnel junctions
embedded in a superconducting loop, threaded by a flux $\Phi_b$, see Fig.~\ref{system}a.
Each Josephson junction is characterized by its critical current $I_0$ and capacitance
$C_0$. Using the mechanical analogy, it can be shown that the dynamics of the SQUID's
phase $\phi$ is isomorphic to that of a fictitious particle of mass $m=2C_0\left(\Phi_0/2\pi\right)^2$
moving in one-dimensional anharmonic potential (harmonic oscillator with weak cubic
perturbation)~\cite{cblaudo04}, see Fig.~\ref{system}b. Here $\Phi_0= h/2e$ denotes the flux quantum. Key
parameters for the potential are the frequency of the bottom of the well $\omega_p$ and
the barrier height $\Delta U$. In the presence of a time-dependent external magnetic flux
$\Phi_b(t)$, the quantum dynamics is described by the total Hamiltonian $\hat
H_\mathrm{tot}=\hat H_\mathrm{\phi}+\hat H_\mathrm{c}$ where
\be \label{eq:AnhrmonicHam}
\hat H_\mathrm{\phi}=\frac{1}{2}\hbar\omega_p\left(\hat P^2+\hat
X^2\right)-\sigma\hbar\omega_p\hat X^3, \quad
H_\mathrm{c}=\hbar\omega_p{\varepsilon(t)}\hat X \,.
\ee
Here $\hat P=\left(1/\sqrt{m\hbar\omega_p}\right)P$ and $\hat X=
\left(\sqrt{m\omega_p/\hbar}\right)\phi$ are the reduced momentum and position operators,
respectively. The anharmonic dimensionless coupling $\sigma$ can be tuned with the bias current; it is small compared to unity. The effect of a
time-dependent external flux $\Phi_b(t)$ is included via the dimensionless function
$\varepsilon(t)=\Phi_b(t)/\Phi_0$.

The theory described by the Hamiltonian $\hat H_\mathrm{\phi}$, the basis Hilbert space
of which is infinite, can be approximated by a theory described by an effective
Hamiltonian (finite $N\times N$ matrix) $ \hat
H_\mathrm{eff}=\sum_{\mu,\nu=0}^{N-1}\bra{\psi_\mu}{\hat H_\mathrm{\phi}}\ket{\psi_\mu}
\ket{\psi_\mu}\bra{\psi_\nu} $ the basis of which is finite. Hamiltonian $\hat
H_\mathrm{eff}$ reproduces only the low-energy physics of the system~\cite{jirari99}.
With the harmonic oscillator eigenfunctions $\ket{\psi_\mu}$ as the expansion basis, the
matrix elements of $\hat H_\mathrm{\phi}$ can be easily calculated. The eigenvalues $E_n$
and wave functions $\ket{n}$ of the effective Hamiltonian can be obtained by
diagonalizing $\hat H_\mathrm{eff}$ using a unitary transformation $\hat
H_\mathrm{eff}=\mathcal{O}^\dagger \hat H_0\mathcal{O}$ where $\hat
H_0=\mbox{diag}(E_0,\ldots,E_{N-1})$. Once the spectrum and wave functions are available, all
physical information can also be obtained, especially the matrix elements of the control
Hamiltonian $\hat H_\mathrm{c}$ in Eq.~(\ref{eq:AnhrmonicHam}). Within the exact
diagonalisation of $\hat H_\mathrm{eff}$, the total Hamiltonian $\hat H_\mathrm{tot}=\hat
H_\mathrm{\phi}+\hat H_\mathrm{c}$ is transformed into the driven N-level quantum system
Hamiltonian whose form is $\hat H=\hat H_\mathrm{0}+\varepsilon(t)\hat H_\mathrm{I}$ with
\begin{align}
\label{eq:InteractionHam}
&\hat H_\mathrm{0}=\sum_{n=0}^{N-1}\,E_n\ket{n}\bra{n};\quad
\hat H_\mathrm{I}=\hbar\omega_p\sum_{n,m=0}^{N-1}d_{n,m}\ket{n}\bra{m}
\end{align}
where $\{\ket{n}: n=0\ldots N-1\}$ is a complete set of orthonormal eigenstates, {\it
i.e.} the eigenvectors of $\hat H_\mathrm{eff}$ corresponding to the energies $E_n$.
Here, the quantities $d_{n,m}$ are the transition dipole moments defined by\break $
d_{n,m}=\langle n|\hat X|m\rangle= \sum_{\mu,\nu}\overline{C}^\mu_n
C^\nu_{m}\int_{-\Lambda}^\Lambda \,dx\overline{\psi}_\mu(x)x\psi_\nu(x) $ where
$C^\mu_{n}=\langle n|\psi_\mu\rangle$ and $\Lambda$ is on the order of
$10\sqrt{{\hbar}/{m\omega_p}}$. The off-diagonal elements of $\hat H_\mathrm{I}$ induce
transitions between energy eigenstates; the diagonal elements renormalize the energy
eigenvalues, an effect known as the Stark shift.


\section{Quantum optimal control problem}

Let time $t$ be in the interval $[t_I=0,t_F]$, for time $t_F$ fixed. An arbitrary state
of the system at time $t$ can be represented by the density matrix $\rho$ acting on
$\mathbb{C}^N$, the Hilbert space of dimension $N$. The density matrix evolves according
to the Liouville-von Neumann equation \be \label{eq:ME} i\hbar\dot{\rho}=\left\lbrack
\hat H,\rho\right\rbrack;\quad \rho(0) =\rho_I \ee where $\rho_I$ is the initial state of
the system. We will use optimal control theory to design a control field
$\varepsilon(t)$ which drives our system from an initial state $\rho_I$ at time $t_I=0$
to a desired target state $\rho_F$ at specified final time $t_F$. The problem can be
formulated in terms of a cost functional that also takes into account experimental
constraints. Minimizing this cost functional leads to the desired physical target, thereby satisfying the constraints.

The question as to whether or not there exists a control that steers the system to a
given goal is of crucial importance. For $N$-level quantum systems subject to a single
control, this question has been addressed in~\cite{ramakrishna95,schirmer01}. In the
absence of dissipative effects, the answer is affirmative because the
dynamical Lie group generated by $i\hat H_\mathrm{0}$ and $i\hat H_\mathrm{I}$ is
isomorphic to the unitary group $U(N)$, which is compact. A target state $\rho_F$ can be
dynamically reached from $\rho_I$ if there exists a unitary operator $\hat U\in U(N)$
such that $\rho_F=\hat U(t_F)\rho_I\hat U(t_F)\dagger$~\cite{schirmer01}. Actually
$U(t_F)$ represents the time evolution operator obeying itself the Schr\"odinger equation
with the initial condition $\hat U(0)=\mathbb{1}$. The formal solution to the
Schr\"odinger equation can be written as $ \hat U(t_F) = {\cal T} \left\{ \exp
\left\lbrack -{i\over\hbar}\int_{0}^{t_F}d\tau\, \hat H \right\rbrack \right\}$, where
the symbol ${\cal T}$ denotes time ordering.


\section{Pontryagin minimum principle}

Suppose the system is prepared at time $t_I=0$ in the initial state $\rho_I$. The
objective is to compute an appropriate time-dependent control function ${\varepsilon}(t)$
steering the system from the initial state $\rho_I$ into a target state $\rho_F$ at fixed
final time $t_F$. The corresponding cost functional may be written as \be
\label{eq:cost} J\lbrack\varepsilon(t)\rbrack= {1 \over 2}\Vert\rho(t_F)-\rho_F\Vert_F^2+ {1 \over
2}\int_0^{t_F}\alpha(t)\varepsilon^2(t)\,dt \ee where $\Vert . \Vert_F$ is the Frobenius
norm: $\Vert A \Vert_F^2=\mbox{Tr}A^\dagger A =\sum_{ij}\left|A_{ij}\right|^2$. Here, the
first term represents the deviation between the state of the system at final time
$\rho(t_F)$ and the target state $\rho_F$, whereas the second integral term penalizes the
field fluency with a generally time-dependent weight $\alpha$. We will illustrate the
physical meaning of $\alpha(t)$ for a specific example below. Minimizing the first term
({\it i.e.} the error) is equivalent to
maximizing the state transfer fidelity $F=\mbox{Tr}\left\{\rho(t_F)\rho_F\right\}$.
Our overall task is to find the control $\varepsilon(t)$ that minimizes
$J\lbrack\varepsilon(t)\rbrack$ and satisfies both the dynamic
constraint and the boundary condition~(\ref{eq:ME}). An optimal solution of this problem
can be obtained using the first order optimality conditions in the form of the Pontryagin
minimum principle (PMP)~\cite{jirari05,sporl05}. These conditions are formulated using a
scalar pseudo Hamiltonian which may in the present case be
cast in the form \be {\mathcal {H}}(\rho,\varepsilon,\lambda) := {1 \over
2}\alpha~\varepsilon^2 + \mbox{Tr} \left\lbrace {\lambda\over i\hbar} \left\lbrack \hat
H,\rho\right\rbrack \right\rbrace, \ee where the adjoint state variable $\lambda$ is an
operator Lagrange multiplier introduced to implement the constraint~(\ref{eq:ME}). The
PMP states that the necessary conditions of optimality for the control problem defined
above are as follows:
\begin{align}
\label{eq:eqrho}
&\dot\rho=\partial_{\lambda}{\mathcal{H}}
=\frac{1}{i\hbar}\left\lbrack \hat H,\rho\right\rbrack,\quad\hat\rho(0)=\rho_I ;\\
\label{eq:eqlambda}
&\dot\lambda =-\partial_{\rho}{\mathcal{H}}
=\frac{1}{i\hbar}\left\lbrack \hat H,\lambda\right\rbrack,\quad\lambda(t_F)=\rho(t_F)-\rho_F;\\
\label{eq:eqepsilon} &0=\partial_{\varepsilon}{\mathcal{H}}=\alpha\varepsilon+
\mbox{Im~Tr} \left\{ \frac{\lambda}{\hbar} \left\lbrack \hat H_\mathrm{I},\rho
\right\rbrack \right\}.
\end{align}
The last condition is also equivalent to a vanishing gradient of the functional $J$
with respect to the control $\varepsilon$,
\begin{align}
&\frac{\delta J}{\delta\varepsilon(t)}
=\mbox{Re}\int_0^{t_F}
{\partial_{\varepsilon}{\mathcal{H}}\left({\rho}(t),
\lambda(t),\varepsilon(t)\right)}\,dt\\
&= \label{eq:gradient} \int_0^{t_F}\left( \alpha(t)\varepsilon(t)+ \mbox{Im~Tr} \left\{
\frac{\lambda(t)}{\hbar} \left\lbrack \hat H_\mathrm{I},\rho(t) \right\rbrack \right\}
\right)\,dt.
\end{align}

Numerically, an iterative procedure based on successive linearization must be employed to
find the optimal control. Here we will use a gradient-based method in order to
find a solution to the system of the necessary conditions of optimality,
Eqs.~(\ref{eq:eqrho})-(\ref{eq:eqepsilon}). More precisely, we have used the L-BFGS-B
routine which is based on a bound constraint quasi-Newton method with BFGS update
rule~\cite{lbfgs}. This routine is appropriate and efficient for solving
constrained as well as unconstrained problems.

\section{Population transfer\label{section:PT}}

We now drive the system from the ground state $\ket{0}$ into one of the
excited states $\ket{n}$, for $n=1,2,3, \ldots$ as an illustrative example of the
efficiency of the control field generated  by the optimal control algorithm.
Suppose the system is in state $\rho_I=\ket{0}\bra{0}$ at time $t=0$. The
objective is to force the system to state $\rho_F=\ket{n}\bra{n}$ for given $n$ at time $t_F$.

A variety of experimental constraints may be imposed in an optimal control problem
in order to select control fields that are feasible from a practical point of
view~\cite{werschnik07}. The purpose of the multiplier $\alpha(t)$ in the cost functional
defined in Eq.~(\ref{eq:cost}) is to force the control field to approach zero at the
initial and final time in accordance with the experiment. For this we use the shape
function~\cite{jirari05}
\be \alpha(t)=\alpha_0+\alpha_1 \left( \exp\left\lbrack-t/\tau\right\rbrack+
\exp\left\lbrack-(t_F-t)/\tau\right\rbrack \right), \ee
where the positive constants $\alpha_j$ are the penalty parameters
and $\tau$ is a rise time. The role of $\alpha_0$ is to penalize
high control field values throughout the time interval $[0,t_F]$;
$\alpha_1$ together with the exponential terms enforces the
field to be nearly zero at the boundaries of the interval
$[0,t_F]$ while simultaneously turning on and off the field
smoothly.

Because all the matrix elements of the interaction Hamiltonian
$H_\mathrm{I}$ defined in Eq.~(\ref{eq:InteractionHam}) are
different from zero the population transfer, for example from
the ground state $\ket{0}$ to the excited state $\ket{4}$, will
involve several pathways. The population transfer can be realized
via a direct transition: $|0\rangle\longrightarrow|4\rangle$ or
via indirect transitions: $|0\rangle\longrightarrow
 |1\rangle\longrightarrow
 |2\rangle\longrightarrow
 |3\rangle\longrightarrow|4\rangle$,~
$|0\rangle\longrightarrow|6\rangle\longrightarrow|4\rangle\ldots$
As a result, the time structure of the control field emerging from
optimal control theory will generally be complicated, due to
quantum mechanical interference between the pathways the field
employs in the process $|0\rangle\longrightarrow|4\rangle$. The
resulting control field is generally characterized by a frequency
content that is not readily implemented experimentally. In order
to reduce the control field complexity, we resort to the spectral
filter technique. A convenient way to restrict the optimal field
to a single desired frequency $\omega_0$ is to filter the gradient
by the formula~\cite{werschnik07,gross91}, \be \label{eq:filter}
\left.\frac{\delta
J}{\delta\varepsilon(t)}\right|_{\mbox{filter}}= \mathcal{F}^{-1}
\left\lbrack g(\omega) \mathcal{F} \left\lbrack \frac{\delta
J}{\delta\varepsilon(t)} \right\rbrack \right\rbrack \ee where \be
\mathcal{F} \left\lbrack \frac{\delta
J}{\delta\varepsilon(t)}\right\rbrack(\omega)
=\frac{1}{\sqrt{2\pi}}\int_{-\infty}^{+\infty} \frac{\delta
J}{\delta\varepsilon(t)}e^{-i\omega t} dt \ee is the Fourier
transform of the gradient of the cost functional defined in
Eq.~(\ref{eq:gradient}) and \be
g(\omega)=e^{-\gamma\left(\omega-\omega_0\right)^2}+
e^{-\gamma\left(\omega+\omega_0\right)^2} \ee is a Gaussian
frequency filter centered around $\pm\omega_0$. Here $\gamma$ is
positive constant, large compared to the unity so that
$g(\omega) =0$ for $\omega\not=\pm\omega_0$ and $g(\omega) =1$ for
$\omega=\pm\omega_0$. During the optimization process, the control
variable is updated by a filtered gradient in each iteration.
Thus, every spectral component in the control variable is
eliminated except the components around $\omega=\pm\omega_0$. The resulting optimized control field has a simple
intuitive interpretation.

The restriction of the spectrum to a single frequency $\omega_0=(E_1-E_0)/\hbar$
simplifies the time structure of the optimal control. Specifically, the time dependence
of the control field can be interpreted as an input signal represented by a single
oscillation of the form \be \varepsilon_{\rm opt}(t)=A(t)\cos\left\lbrack\omega_0 t
+\varphi(t)\right\rbrack \ee where the amplitude $A(t)$ and phase $\varphi(t)$ vary slowly with
time compared to $\omega_0$. The goal of complex demodulation is to extract the amplitude
and phase as function of time~\cite{bloomfield}. First, the original signal
$\varepsilon_{\rm opt}(t)$ is multiplied by a complex modulation of frequency $\omega_0$,
yielding $ f(t)=\varepsilon_{\rm opt}(t)e^{-i\omega_0 t}=
A(t)[e^{i\varphi(t)}+e^{-2i\omega_0 t - i\varphi(t)}]/2$. Passing the
resulting signal $f(t)$ trough an ideal low-pass filter of cutoff frequency $\omega_c <
\omega_0$ leads to $ g(t)= A(t)e^{i\varphi(t)}/2=\alpha(t)+i\beta(t) $. The time-dependent
amplitude and phase can then be calculated as $A(t)=2\sqrt{\alpha(t)^2+\beta(t)^2}$ and
$\varphi(t)=\arctan\left\lbrack\beta(t)/\alpha(t)\right\rbrack$, respectively.

\section{Results}
We now illustrate the above concepts with the aid of some representative
examples. Throughout this section, for the numerical simulations of population transfer
from the ground state $|0\rangle$ to one of the excited states $|n\rangle$, we use the
following nominal parameters, typical for experiments with current-biased
SQUIDs~\cite{cblaudo04} : anharmonicity $\sigma=0.0325$; target time
$t_F=500/\omega_p$, chosen to be short enough to avoid substantial relaxation and
decoherence phenomena; time step $\Delta t= 3.0\times 10^{-2}/\omega_p$ corresponding to $M=2^{14}$
as the number of mesh points; frequency filter $\omega_0=(E_1-E_0)/\hbar$; penalties
factors $\alpha_0=10^{-1}$; $\alpha_1=10^{2}$ and rise time $\tau= 10^2
/\omega_p$.

\begin{figure}[!t]
\centerline{\includegraphics[height=1\columnwidth,angle=-90]{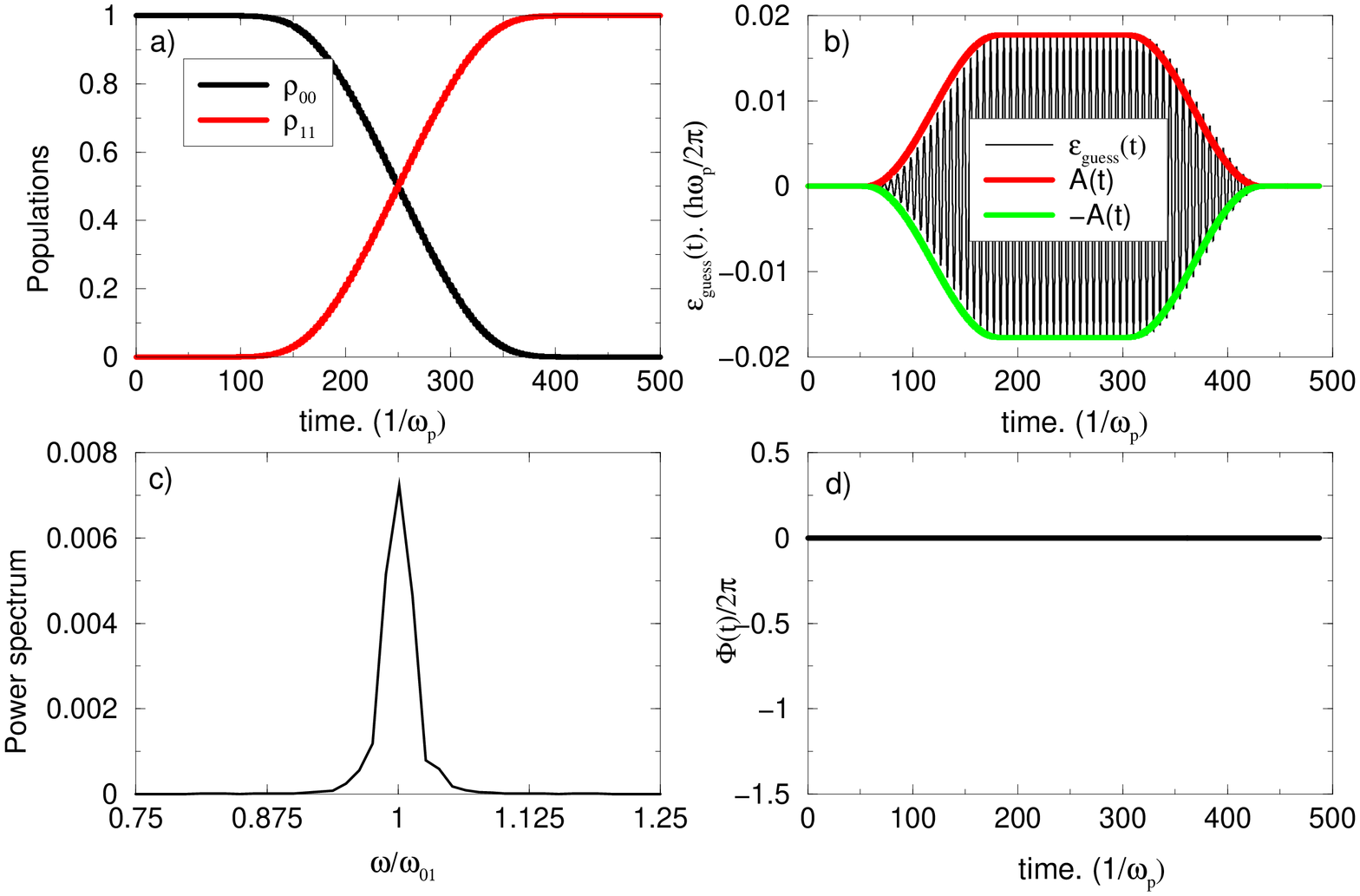}}
\caption{Population transfer from the ground state $|0\rangle$ to the excited state
$|1\rangle$, following a $\pi$-pulse. Panel (a) shows the numerically obtained evolution
of populations, (b) and (c), respectively, show the $\pi$-pulse and its numerically
obtained power spectrum. The amplitude of the pulse versus time is also shown in panel
(b) while the numerically obtained time evolution of the phase is displayed in the panel
(d).} \label{fig:fig1}
\centerline{\includegraphics[height=1\columnwidth,angle=-90]{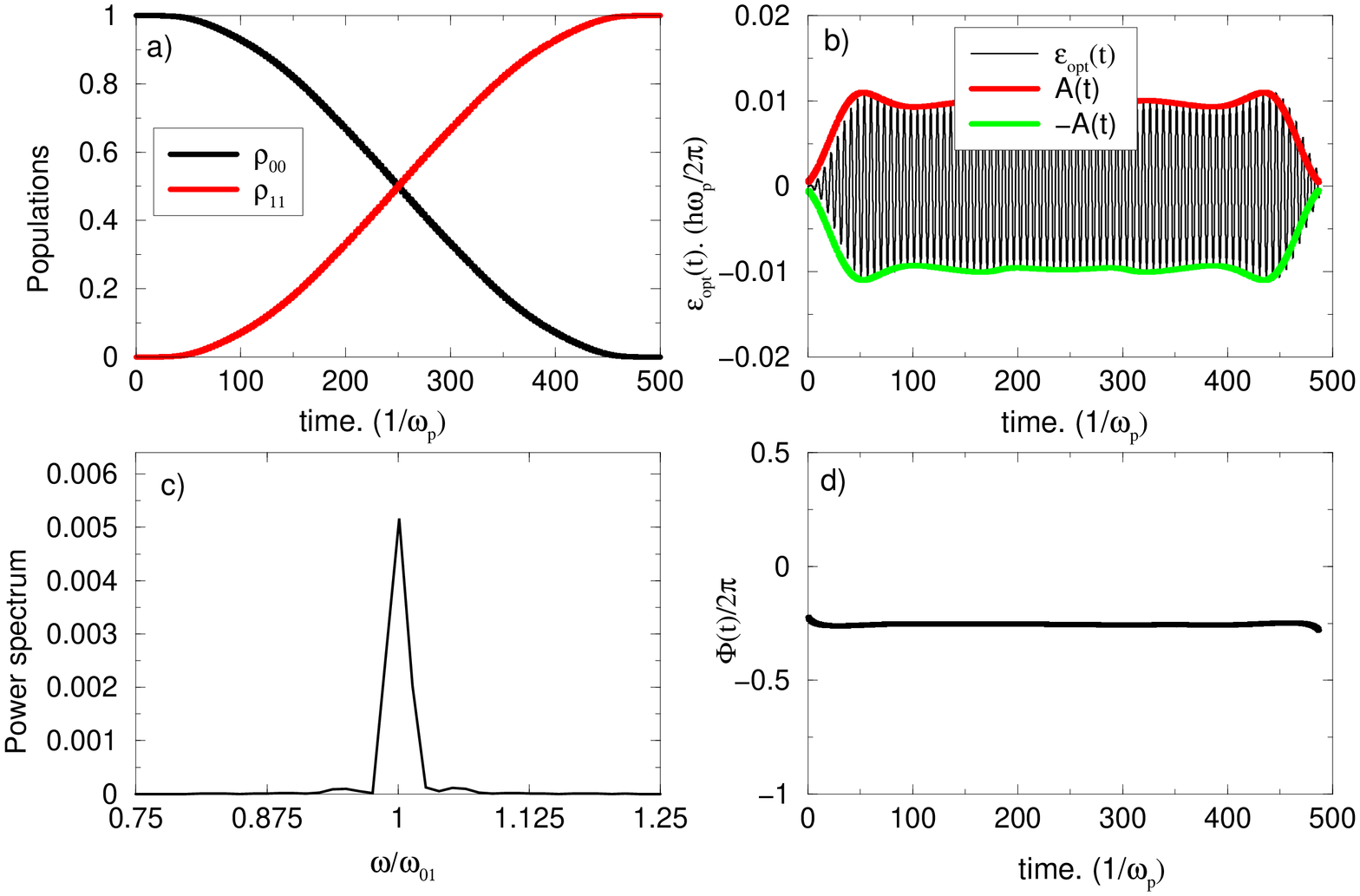}} \caption{Optimal
control of population transfer from the ground state $|0\rangle$ to the excited state
$|1\rangle$: (a) shows the evolution of populations, (b) and (c), respectively, show the
selected control field and its power spectrum. The amplitude of the control field versus
time is also shown in the panel (b) while the time evolution of the phase is displayed in
the panel (d).} \label{fig:fig2}
\end{figure}

We start by considering the population transfer in a two-level system, $N=2$. It
is well-known that population transfer can be obtained within the rotating wave
approximation (RWA) by applying a so-called $\pi$-pulse: a time-dependent
signal of frequency $\omega_0 = E_1 - E_0$ ({\it i.e.} resonant with the level spacing) with
duration equal to half of the so-called Rabi period. It is useful to study this simple example
with the help of optimal control theory, as it enables one to compare to the known result
for a $\pi$-pulse and hence test the numerical implementation; it also sheds some
light on the functioning of the control procedure.

\begin{figure}[!t]
\centerline{\includegraphics[height=1\columnwidth,angle=-90]{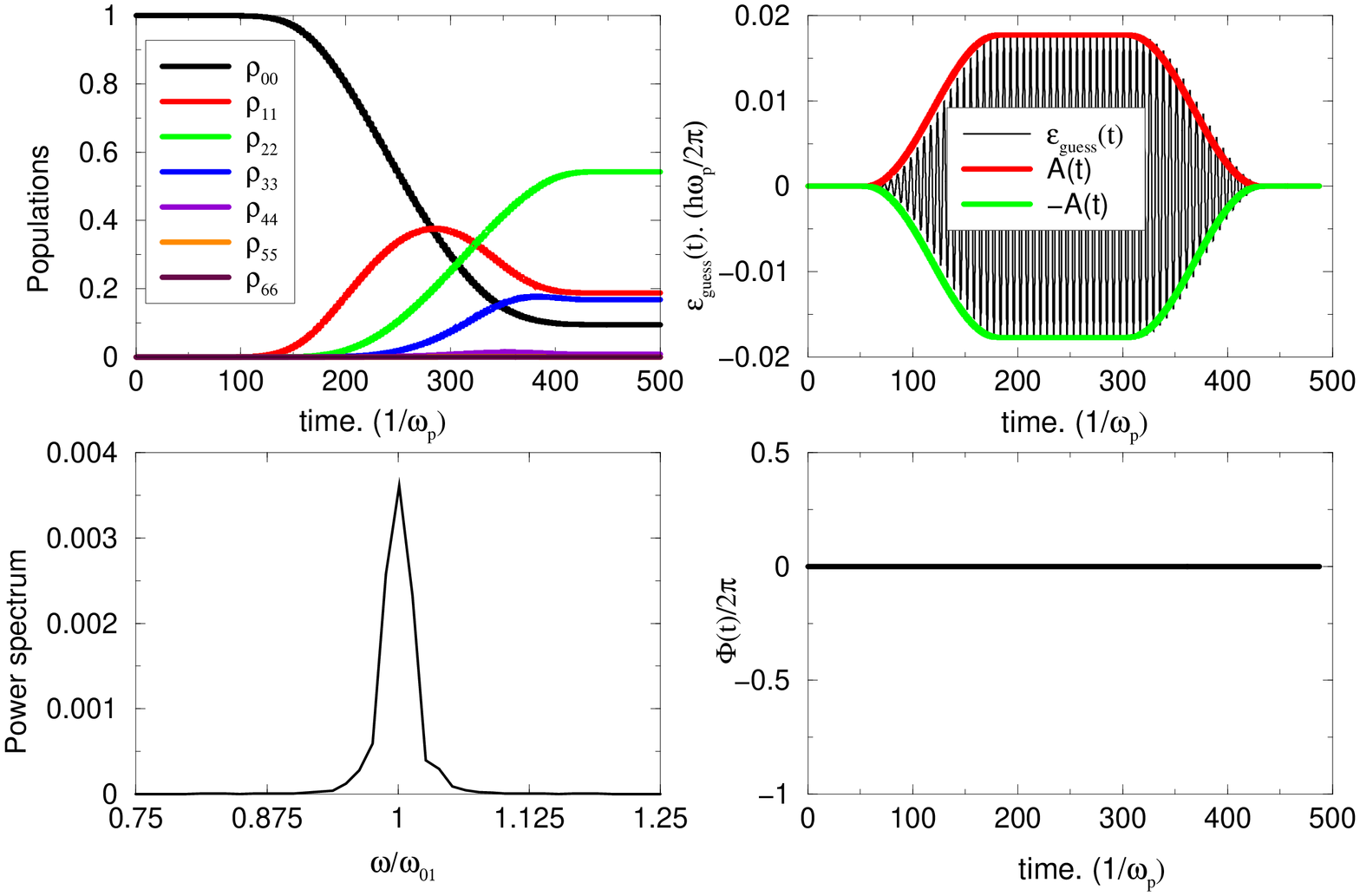}}
\caption{Population transfer from the ground state $|0\rangle$ to the excited state
$|1\rangle$ in a seven-level system, using a $\pi$-pulse; (a), (b), (c) and (d), as in the previous Figure.} \label{fig:fig3}
\centerline{\includegraphics[height=1\columnwidth,angle=-90]{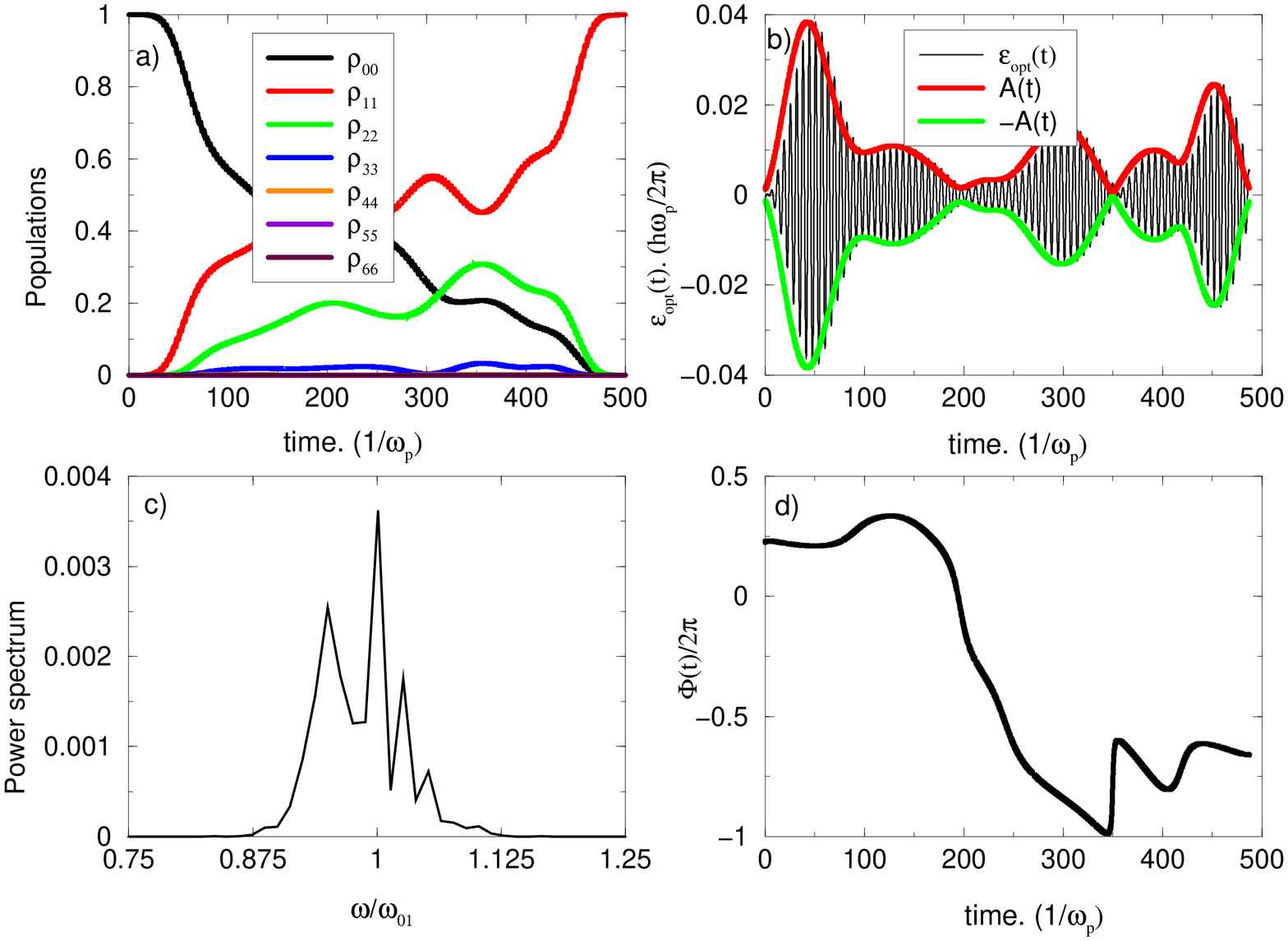}} \caption{Control
of population transfer from the ground state $|0\rangle$ to the excited state $|1\rangle$
in a seven-level system, using the $\pi$-pulse as a gues; (a), (b), (c) and (d), as in the previous Figure.} \label{fig:fig4}
\end{figure}

We first study the response of a two-level system to a $\pi$-pulse. The pulse
is shown Fig.~\ref{fig:fig1}b, its complex demodulation is shown
in panels (c) and (d). The response of the two-level system is obtained numerically and shown in
Fig.~\ref{fig:fig1}a; we see that perfect population transfer is achieved. We
subsequently use the $\pi$-pulse of Fig.~\ref{fig:fig1}b as a guess for an optimal
control simulation, the results of which are shown in Fig.~\ref{fig:fig2}. It can be seen
that the application of optimal control slightly modifies the original $\pi$-pulse.
Specifically, the overall amplitude is smaller as a result of the minimal amplitude constraint we imposed; the pulse duration has increased
in order to conserve the total area of the envelope and hence the
$\pi$ nature of the pulse.

We next wish to implement a population transfer from the ground state $|0\rangle$ to the
excited state $|1\rangle$ in a multilevel system with $N=6$. As an initial guess we use
the $\pi$-pulse of Fig.~\ref{fig:fig1}b. The response of the seven-level system to this
guess is shown in Fig.~\ref{fig:fig3}a: the presence of the levels beyond $n=1$ clearly
leads to strong contamination effects that limit the efficiency of the transfer. We then
use the initial guess as a starting point for an optimal control simulation; the
resulting optimal pulse is shown in Fig.~\ref{fig:fig4}b, the result of the corresponding
population transfer in panel (a). In order to avoid contamination, both the pulse amplitude
and its phase vary smoothly as a function of time, as shown in panels (c) and (d). We
verified that a transfer from $|0\rangle$ to $|1\rangle$ can also be achieved in principle
with a $\pi$-pulse similar to the one shown in Fig.~4b. However, the overall amplitude of the required pulse should be
much smaller (by about a factor ten) in order to avoid non-resonant
transitions to the higher levels. Accordingly the resulting duration would be ten times
longer, thereby exceeding typical relaxation and decoherence times.

\begin{figure}
\centerline{\includegraphics[height=1\columnwidth,angle=-90]{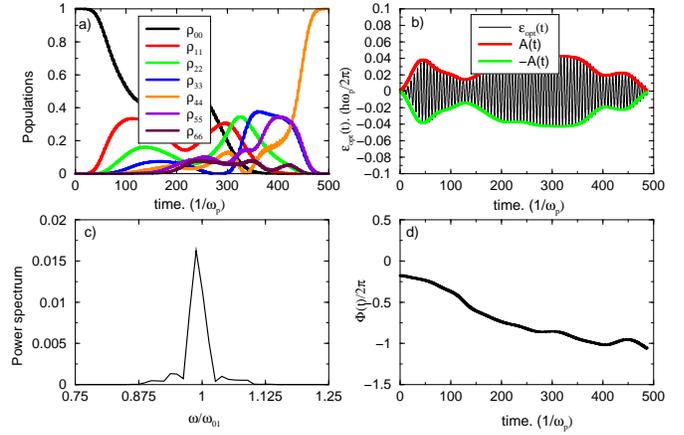}} \caption{Control of
opulation transfer from the ground state $|0\rangle$ to the excited state $|4\rangle$ in
a seven-level system; (a), (b), (c) and (d), as in the previous Figure.} \label{fig:fig5}
\end{figure}

Finally, in Fig.~\ref{fig:fig5}, we show an optimal control simulation for the population
transfer from the ground state $|0\rangle$ to the excited state $|4\rangle$. As can be
seen by comparing panel (a) of Fig.~\ref{fig:fig5} with that of Fig.~\ref{fig:fig4}, this
transfer involves enhanced occupation during manipulation of the excited states of the
system. We therefore expect transfers to higher states to be less robust against noise
than transfers to low-lying excited states (see also the next Section).

\section{Sensitivity analysis}

In applications, the system parameters are usually not fixed but may be subject
to perturbations and noises. For instance, the environment of the dc-SDUID induces
time-dependent fluctuations of the bias current and flux. Because
of the external perturbations, practical devices are not capable of operating precisely
neither at the prescribed system parameters nor at the computed control field. Then, it
is of great importance to know the sensitivity of the optimal solution
with respect to perturbations of any of the system parameters. In this section, we shall
only consider the influence of slow, adiabatic fluctuations. In the
presence of this so-called adiabatic noise, the system parameters remain constant during
a given manipulation but fluctuate during repetitive measurements needed to obtain
quantum statistics.

\begin{figure}[!t]
\centerline{\includegraphics[height=1.0\columnwidth,angle=-90]{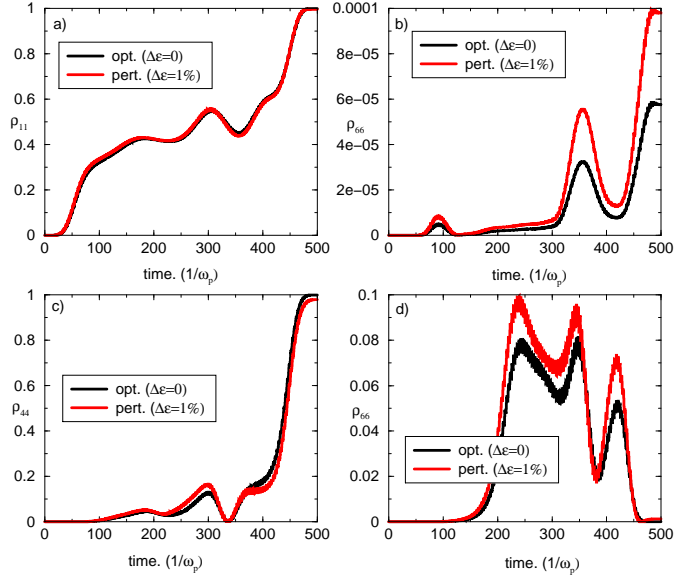}}
\caption{Sensitivity of the population transfer to a small amplitude
perturbation of the optimal control field.}
\label{fig:fig6}
\end{figure}

In order to develop some feeling for the sensitivity of the model system discussed here with respect to adiabatic noise, we show in
Fig.~\ref{fig:fig6} the effect of a small static amplitude fluctuation $\Delta \epsilon$ (of
about 1\%) with respect to the optimal amplitude on the result of the optimal control of a seven-level system when a population
transfer from $|0\rangle$ to $|1\rangle$ is sought [panel (a) and (b)] or when a transfer
from $|0\rangle$ to $|4\rangle$ is sought [panel (c) and (d)]. As is to be expected, the
sensitivity to such a fluctuation is larger in the latter case, which involves more
excited states during transfer.

\begin{figure}[!t]
\centerline{\includegraphics[width=0.8\columnwidth,angle=0]{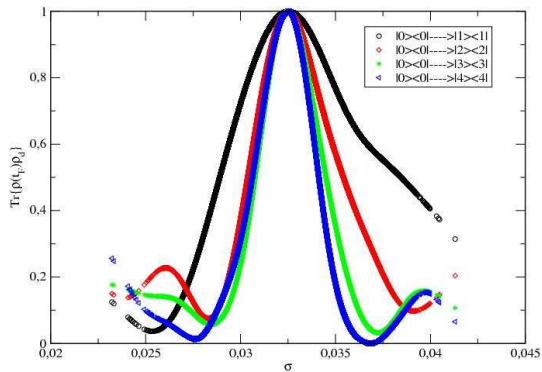}}
\caption{Sensitivity of the population transfer fidelity to the random fluctuations of
the anharmonic coupling.} \label{fig:fig7}
\end{figure}

We finally consider the effect of a slow variation of a system parameter during
application of an optimal control pulse. Specifically, we assume variations of the anharmonic coupling
$\sigma$ to occur when the same optimal control sequence is repeated many times. Because the anharmonic
coupling depends on the so-called working point, every random perturbation
that changes the bias current or flux changes the parameter
$\sigma$ leading to uncertainties in the eigenvalues $E_n$ and the transition dipole
moments $d_{n,m}$ in Eq.~(\ref{eq:InteractionHam}). In Fig.~\ref{fig:fig7} we show a plot of the state transfer fidelity $F$ as a function of $\sigma$, calculated for a control field optimized for the case $\sigma = \overline\sigma =0.0325$. The fidelity drops as $\sigma$ deviates from $\overline\sigma$; we see again that it decreases faster as transitions to higher levels are considered. From this result we can calculate the average fidelity $\overline{\mbox{F}}= {1\over M}\sum_{i=1}^M F_i= {1\over
M}\sum_{i=1}^M\mbox{Tr}\left\{\rho_i(t_F)\rho_d\right\} $, when the optimized control sequence is repeated $M$ times, assuming $\sigma$ to be a normally distributed random variable with a mean $\overline\sigma=0.0325$ and a standard
deviation $\Delta\sigma=\overline\sigma/16$. This choice corresponds to a typical experimental low-frequency noise present in phase qubits~\cite{neeley08,hoskinson08}, yielding a Q-factor of about $1000$. We find the following average state transfer fidelities: $\overline{\mbox{F}}_{|0\rangle \to |1\rangle} = 85\%$, $\overline{\mbox{F}}_{|0\rangle \to |2\rangle} = 73\%$, $\overline{\mbox{F}}_{|0\rangle \to |3\rangle} = 60\%$, and $\overline{\mbox{F}}_{|0\rangle \to |4\rangle} = 55\%$. If the low-frequency noise will be reduced only by a factor five (Q-factors $\sim 5000$), all the above fidelities remain larger than $95\%$.

\section{Conclusion}
Using optimal control theory, we have studied the possibility of a
population transfer from the ground state to an arbitrary excited
state of a superconducting quantum N-level system. We have found that such state transfer
can be obtained with good fidelity, using optimized pulses which can be realized using existing microwave technology.
We have considered the effects of low-frequency noise and found that it reduces the average state transfer fidelity. Using parameters describing actual phase qubits, we find a loss of fidelity of about 15 $\%$ for the transfer from level $|0\rangle$ to $|1\rangle$ up to 45 $\%$, for a transfer from level $|0\rangle$ to $|4\rangle$. However, a substantial improvement of fidelity is possible by slightly decreasing the effects of low-frequency noise.

We thank R. Fazio and F. Taddei for useful discussions. Financial support from the European network EUROSQIP is gratefully acknowledged.


\begin{thebibliography}{99}

\bibitem{makhlin01}
\Name{Yu. Makhlin, G. Sch\"on, and S. Shnirman}
\Book{Rev. Mod. Phys. 73, 357-400 (2001)}.

\bibitem{wendin05}
\Name{G. Wendin \and V. S. Shumeiko}
\Book{Superconducting quantum circuits, qubits and computing,cond-mat/0508729}.

\bibitem{martinis02}
\Name{J. M. Martinis, S. Nam, J. Aumentado, \and C. Urbina}
\REVIEW{Phys. Rev. Lett.}{89}{2002}{117901}.

\bibitem{neeley08}
\Name{M. Neeley {\em et al.}}
\REVIEW{Nature Physics}{4}{2008}{523}.

\bibitem{hoskinson08}
\Name{E. Hoskinson {\em et al.}}
\Book{arXiv: 0811.2174}.

\bibitem{cblaudo04}
\Name{J. Claudon, F. Balestro, F. W. J. Hekking \and O. Buisson}
\REVIEW{Phys. Rev. Lett.}{93}{2004}{187003}.

\bibitem{fazio99}
\Name{R. Fazio, G. M. Palma \and J. Siewert}
\REVIEW{Phys. Rev. Lett.}{83}{1999}{5385}.

\bibitem{genovese07}
\Name{Marco Genovese \and Paolo Traina}
\Book{Review on qudits production and their application to Quantum Communication
and Studies on Local Realism,
quant-ph/07111288}.

\bibitem{ARTHURE}
\Name{Bryson A. E. \and Ho Y. C}
\Book{Applied Optimal Control}
\Editor{Hemisphere, New York}
\Year{1975}.

\bibitem{shi88}
\Name{S. Shi \and H. Rabitz}
\REVIEW{J. Chem. Phys.}{92}{1988}{364}{1988}.

\bibitem{grace07}
\Name{M. Grace, C. Brif, H. Rabitz, I. A. Walmsley, R. L. Kosut \and D. A. Lidar}
\REVIEW{J. Phys. B: At. Mol. Opt. Phys.}{40}{2007}{S103}.

\bibitem{safaei_08}
\Name{S.Safaei, S. Montangero, F.Taddei \and R. Fazio}
\Book{arXiv: 0811.2174}.

\bibitem{gross91}
\Name{P. Gross, D. Neuhauser \and H. Rabitz}
\REVIEW{J. Chem. Phys.}{96}{1991}{2834}.

\bibitem{werschnik07}
\Name{J. Werschnik and E. K. U. Gross}
\REVIEW{J. Phys. B: At. Mol. Opti. Phys.}{40}{2007}{R175}.

\bibitem{jirari99}
\Name{H. Jirari, H. Kr\"oger, X.Q. Luo, and K. Moriarty}
\REVIEW{Phys. Lett. A}{258}{1999}{6}.

\bibitem{ramakrishna95}
\Name{V. Ramakrishna, M. V. Salapaka, M. Dahleh, H. Rabitz \and A. Pierce}
\REVIEW{Phys. Rev. A.}{51}{1995}{960}.

\bibitem{schirmer01}
\Name{S. G. Schirmer, H. Fu \and A. I. Solomon}
\REVIEW{Phys. Rev. A.}{63}{2001}{063410}.

\bibitem{jirari05}
\Name{H. Jirari \and W. P\"otz}
\REVIEW{Phys. Rev. A}{72}{2005}{013409};
\REVIEW{Phys. Rev. A}{74}{2006}{022306};
\REVIEW{Euro. Phys. Lett.}{77}{2007}{50005}.

\bibitem{sporl05}
\Name{A. K. Sp\"orl, T. Schulte-Herbr\"uggen, S. J. Glaser,
V. Bergholm, M. J. Storcz, J. Ferber, \and F. K. Wilhelm}
\REVIEW{Phys. Rev. A}{75}{2007}{012302}.

\bibitem{lbfgs}
\Book{http://neos.mcs.anl.gov/neos/solvers/bco:L-BFGS-B/}.

\bibitem{bloomfield}
\Name{P. Bloomfield}
\Book{Fourier Analysis of time series: an introduction}
\Editor{Wiley, New York}
\Year{1976}
\Page{147}.





\end{thebibliography}
\end{document}